\documentclass{ws-procs9x6}

\usepackage{amsmath,amssymb,graphicx,color}

\newcommand{\MQSS}{M_{\text{QSS}}}

\begin{document}

\title{OUT--OF--EQUILIBRIUM PHASE TRANSITIONS \\ IN MEAN--FIELD HAMILTONIAN DYNAMICS }

\author{PIERRE-HENRI CHAVANIS}

\address{Laboratoire de Physique Th\'eorique, Universit\'e Paul Sabatier, 118,
route de Narbonne 31062 Toulouse, France \\
E-mail: chavanis@irsamc.ups-tlse.fr}

\author{GIOVANNI DE NINNO}

\address{Sincrotrone Trieste, S.S. 14 Km 163.5, Basovizza, 34012, Trieste, Italy \\
University of Nova Gorica, Vipavska 13, POB 301, SI-5000, Nova Gorica, Slovenia \\
E-mail: giovanni.deninno@elettra.trieste.it}

\author{DUCCIO FANELLI}

\address{Theoretical Physics, School of Physics and Astronomy, University of Manchester, 
Manchester M13 9PL, United Kingdom \\ 
E-mail: Duccio.Fanelli@manchester.ac.uk}

\author{STEFANO RUFFO}

\address{Dipartimento di Energetica ``S. Stecco'' and CSDC, Universit\`a di
Firenze, and INFN, via S. Marta, 3, 50139 Firenze, Italy \\
E-mail: stefano.ruffo@unifi.it}

\begin{abstract}
Systems with long-range interactions display a short-time relaxation
towards Quasi--Stationary States (QSSs), whose lifetime increases with
system size. With reference to the Hamiltonian Mean Field (HMF) model,
we here review Lynden-Bell's theory of ``violent relaxation''. The
latter results in a maximum entropy scheme for a water-bag initial
profile which predicts the presence of {\it out--of--equilibrium phase
transitions} separating homogeneous (zero magnetization)
from inhomogeneous (non--zero magnetization) QSSs. 
Two different parametric representations of the initial condition are analyzed
and the features of the phase diagram are discussed. In both representations we
find a second order and a first order line of phase transitions that merge at
a tricritical point. Particular attention is payed to the
condition of existence and stability of the homogenous phase.
\end{abstract}

\keywords{Quasi-stationary states, Hamiltonian Mean-Field model, 
Out--of--equilibrium phase transitions}

\bodymatter

\section{Introduction}
\label{s:intro}

Hamiltonian systems arise in many branches of applied and fundamental
physics and, in this respect, constitute a universal framework of
extraordinary conceptual importance. Spectacular
examples are undoubtely found in the astrophysical
context. The process of hierarchical clustering via
gravitational instability, which gives birth to the galaxies~\cite{astro}, can in fact be cast in a Hamiltonian
setting. Surprisingly enough, the galaxies that we observe have not
yet relaxed to thermodynamic equilibrium and possibly correspond to
intermediate Quasi--Stationary States (QSSs). The latter are
in a long-lasting dynamical regime, whose lifetime diverges with the size of the
system. The emergence of such states has been reported in several
different domains, ranging from charged cold plasmas
\cite{turchetti} to Free Electron Lasers (FELs) \cite{barre}, and
long-range forces have been hypothesized to be intimately connected to
those peculiar phenomena.

Long-range interactions are such that the two-body interaction
potential decays at large distances with a power--law exponent which
is smaller than the space dimension. The dynamical and thermodynamical
properties of physical systems subject to long-range couplings were
poorly understood until a few years ago, and their study was
essentially restricted to astrophysics (e.g., self-gravitating
systems). Later, it was recognized that long--range systems exhibit
universal, albeit unconventional, equilibrium and out--of--equilibrium
features \cite{Houches02}. Besides slow relaxation to equilibrium,
these include ensemble inequivalence (negative specific heat,
temperature jumps), violations of ergodicity and disconnection of the
energy surface, subtleties in the relation of the fluid
(i.e. continuum) picture and the particle (granular) picture, new
macroscopic quantum effects, etc..

While progress has been made in understanding such phenomena, an
overall interpretative framework is, however, still lacking. In
particular, even though the ubiquity of QSSs has been accepted as an
important general concept in non-equilibrium statistical mechanics,
different, contrasting, attempts to explain their emergence have
catalysed a vigorous discussion in the literature \cite{debate}.

To shed light onto this fascinating field, one can resort to toy
models which have the merit of capturing basic physical modalities,
while allowing for a dramatic reduction in complexity. This is the
case of the so-called Hamiltonian Mean Field (HMF) model which
describes the evolution of N rotators, coupled through an equal
strength, attractive or repulsive, cosine
interaction~\cite{antoni-95}. The Hamiltonian, in the attractive case, reads
\begin{equation}
\label{eq:ham}
H = \frac{1}{2} \sum_{j=1}^N p_j^2 + \frac{1}{2 N} \sum_{i,j=1}^N [1-\cos(\theta_j-\theta_i)]~,
\end{equation}
where $\theta_j$ represents the orientation of the $j$-th rotator and
$p_j$ stands for the conjugated momentum. To monitor the evolution of
the system, it is customary to introduce the magnetization, an order
parameter defined as $M=|{\mathbf M}|=|\sum {\mathbf m_i}| /N$, where
${\mathbf m_i}=(\cos \theta_i,\sin \theta_i)$ is the magnetization
vector. The HMF model shares many similarities with gravitational and
charged sheet models~\cite{Konishi,ElskensEscande} and has been
extensively studied \cite{chavanis} as a paradigmatic representative
of the broad class of systems with long-range interactions.  The
equilibrium solution is straightforwardly worked out~\cite{antoni-95}
and reveals the existence of a second-order phase transition at the
critical energy density $U_{c}=3/4$: below this threshold value the
Boltzmann-Gibbs equilibrium state is inhomogeneous (magnetized).

In the following, we shall discuss the appearence of QSSs in the HMF
setting and review a maximum entropy principle aimed at explaining the
behaviour of out-of-equilibrium macroscopic observables. The proposed
approach is founded on the observation that in the continuum limit
(for an infinite number of particles) the discrete HMF equations
converge towards the Vlasov equation, which governs the evolution of
the single--particle distribution function (DF). Within this scenario,
the QSSs correspond to statistical equilibria of the continuous Vlasov
model. As we shall see, the theory allows us to accurately predict
out-of-equilibrium phase transitions separating the homogeneous
(non-magnetized) and inhomogeneous (magnetized) phases
\cite{chavanis1,antoniazzi-06}.  Special attention is here
devoted to characterizing analytically the
basin of existence of the homogeneous zone. Concerning the structure of the phase diagram,
a bridge between the two possible formal settings, respectively
\cite{chavanis1} and \cite{yama}, is here established.

The paper is organized as follows. In Section~\ref{s:QSS_1}
we present the continuous Vlasov picture and discuss the maximum entropy scheme. The
properties of the homogeneous solution are
highlighted in Section~\ref{homogeneous}, where conditions of existence are also
derived. Section~\ref{stability} is devoted to analyze the stability
of the homogeneous phase. A detailed account of the phase diagram is
provided in Sections \ref{diagr1} and \ref{diagr2}, where the case of
a ``rectangular'' and generic water--bag initial distribution are
respectively considered. Finally, in Section~\ref{concl} we sum up and
draw our conclusions.

\section{On the emergence of quasi-stationary states: Predictions from the Lynden-Bell theory 
within the Vlasov picture}
\label{s:QSS_1}

As previously mentioned, long-range systems can be trapped in
long-lasting Quasi-Stationary-States (QSSs)~\cite{Latora}, before
relaxing to Boltzmann thermal equilibrium. The
existence of QSSs was firstly recognized with reference to
galactic and cosmological applications
(see~\cite{Konishi} and references therein) and then, more recently,
re-discovered in other fields, e.g.two-dimensional turbulence
\cite{hd} and plasma-wave interactions \cite{ElskensEscande}. Interestingly, 
when performing the infinite size limit $N \rightarrow \infty$ {\it
before} the infinite time limit, $t \rightarrow \infty$, the system
remains indefinitely confined in the QSSs \cite{tsallis}. For this
reason, QSSs are expected to play a relevant role in systems composed
by a large number of particles subject to long-range couplings, where
they are likely to constitute the solely experimentally accessible
dynamical regimes \cite{barre, turchetti}.

\begin{figure}[htbp]
  \centering \vspace*{2em} \includegraphics[width=10cm]{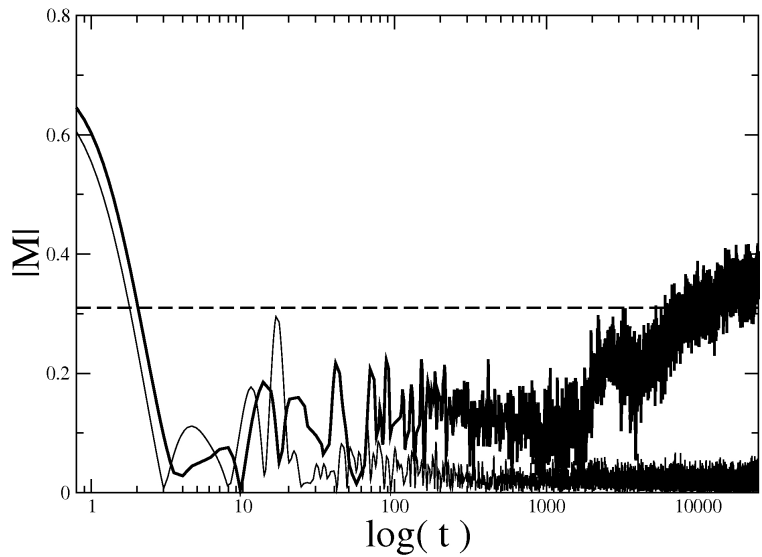}
  \caption{Magnetization $M(t)$ as function of time $t$. In both
  cases, an initial ``violent" relaxation toward the QSS regime is
  displayed. The time series relative to $N=1000$ (thick full line) 
  converges more rapidly to the Boltzmann equilibrium
  solution (dashed horizontal line). When the number of simulated
  particles is increased, $N=10000$ (thin full line), the relaxation to
  equilibrium gets slower (the convergence towards the
  Boltzmann plateau is outside the frame of the figure). Simulations
  are carried on for a rectangular water--bag initial distribution,
  see Eq.~(\ref{water--bag_rect}). 
  } \label{fig:QSS}
\end{figure}

QSSs are also found in the HMF model, as clearly testified in
Fig.~\ref{fig:QSS}. Here, the magnetization is monitored as a function
of time, for two different values of $N$. The larger the system the
longer the intermediate phase where it remains confined before
reaching the final equilibrium. In a recent series of papers
\cite{chavanis1, barre, antoniazzi-06, califano, yama}, an approximate
analytical theory based on the Vlasov equation has been proposed which
stems from the seminal work of Lynden-Bell~\cite{LyndenBell67}. This
is a fully predictive approach, justified from first
principles, which captures most of the peculiar
traits of the HMF out--of--equilibrium dynamics. The
philosophy of the proposed approach, as well as the main predictions
derived within this framework, are reviewed in the following.

In the limit of $N \rightarrow \infty$, the HMF system can be formally replaced by 
the following Vlasov equation
\begin{equation}
\frac{\partial f}{\partial t} + p\frac{\partial f}
{\partial \theta} - \left( M_x[f] \sin \theta - M_y[f]\cos \theta \right) \frac{\partial f}{\partial p}=0,
\label{eq:VlasovHMF}
\end{equation}
where $f(\theta,p,t)$ is the one-body microscopic distribution function normalized such 
that ${\cal M}[f] \equiv \int f d\theta dp=1$, and the two components of the complex magnetization 
are respectively given by 

\begin{eqnarray}
\label{magnetization}
M_x[f]&=&\int f \cos \theta d\theta dp, \\
M_y[f]&=&\int f \sin \theta d\theta dp. \nonumber
\end{eqnarray}

The mean field energy can be expressed as
\begin{eqnarray}
\label{energy}
U=\frac{1}{2}\int f p^{2}d\theta dp-\frac{M_x^2+M_y^2}{2}+\frac{1}{2}.
\end{eqnarray}

\begin{figure}[htbp]
  \centering \vspace*{2em} \includegraphics[width=10cm]{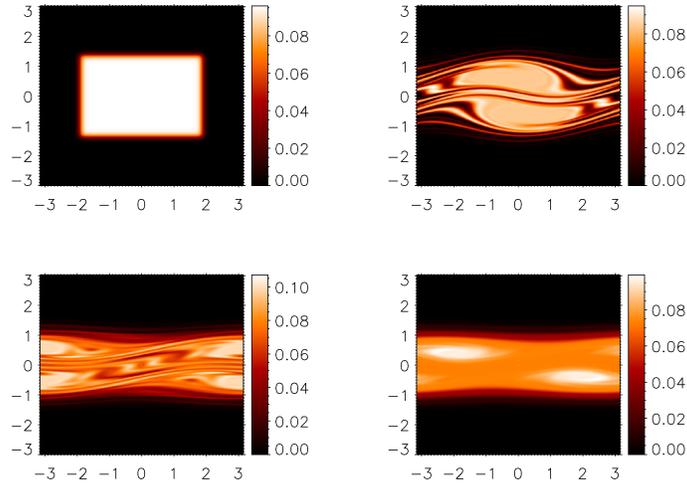}
  \caption{The process of phase mixing is here illustrated, showing
  four snapshots of the time evolution of an initial rectangular water--bag
  distribution. The final state (right bottom) is a QSS.} \label{fig:rectangular}
\end{figure}

Working in this setting, it can be then hypothesized that QSSs
correspond to stationary equilibria of the Vlasov equation and hence
resort to the pioneering Lynden-Bell's violent relaxation
theory~\cite{LyndenBell67} . The latter was initially devised to
investigate the process of galaxy formation via gravitational
instability and later on applied to the two-dimensional Euler
equation~\cite{Chavanis96}. The main idea goes as follows. The Vlasov
dynamics induces a progressive filamentation of the initial single
particle distribution profile, i.e. the continuous counterpart of the
discrete N-body distribution, which proceeds at smaller and smaller
scales without reaching an equilibrium. Conversely, at a coarse
grained level the process comes to an end, and the distribution
function $\bar{f}(\theta,p,t)$, averaged over a finite grid,
eventually converges to an asymptotic form. 
The time evolution of a rectangular water--bag initial distribution is
shown in Fig.~\ref{fig:rectangular}

Following Lynden-Bell, one can associate a mixing entropy to
this process and calculate the statistical
equilibrium by maximizing the entropy, while imposing the conservation
of the Vlasov dynamical invariants. More specifically, the above
procedure implicitly requires that the system mixes well, in which
case, assuming ergodicity (efficient mixing), the QSS predicted by
Lynden-Bell, $\bar{f}_{QSS}(\theta,p,t)$, is obtained by maximizing
the mixing entropy. As a side remark, it is also worth emphasising that
{\it the prediction of the QSS depends on the details of the
initial condition} \cite{super}, not only on the value of the mass
${\cal M}$ and energy $U$ as for the Boltzmann statistical equilibrium state. 
This is due to the fact that the Vlasov equation admits an infinite number of invariants, the
Casimirs or, equivalently, the moments ${\cal M}_{n}=\int \overline{f^{n}}d\theta dp$ 
of the fine-grained distribution function. In the following, we shall consider a 
{\it very simple} initial condition where the distribution function takes
only two values $f_{0}$ and $0$. In that case, the invariants reduce to ${\cal M}$ and $U$ 
since the moments ${\cal M}_{n>1}$ can all be expressed in terms of  ${\cal M}$ and $f_{0}$ 
as ${\cal M}_{n}=\int \overline{f^{n}}d\theta dp=\int \overline{f_{0}^{n-1}
\times f}d\theta dp=\int f_0^{n-1} \overline{f} d\theta dp=f_0^{n-1} {\cal M}$. 
For the specific case at hand, the Lynden-Bell entropy is then explicitly constructed from the
coarse-grained DF $\bar{f}$ and reads

\begin{equation}
S[\bar{f}]=-\int \!\!{\mathrm d}p{\mathrm d}\theta \, 
\left[\frac{\bar{f}}{f_0} \ln \frac{\bar{f}}{f_0}
+\left(1-\frac{\bar{f}}{f_0}\right)\ln
\left(1-\frac{\bar{f}}{f_0}\right)\right].
\label{eq:entropieVlasov}
\end{equation}

We thus have to solve the optimization problem
\footnote{The momentum $P=\int f p d\theta dp$ is also
a conserved quantity but since we look for solutions where the total
momentum is zero, the corresponding Lagrange multiplier $\gamma$
vanishes trivially \cite{antoniazzi-06} so that, for convenience, we
can ignore this constraint right from the begining.}
\begin{eqnarray}
\label{vr7}
\max_{\overline{f}}\lbrace S[\overline{f}]\quad |\quad U[\overline{f}]=U, 
{\cal M} [\overline{f}]=1 \rbrace.
\end{eqnarray} 
The maximization problem (\ref{vr7}) is also a condition
of formal nonlinear dynamical stability with respect to the Vlasov
equation, according to the refined stability criterion of Ellis {\it
et al.} \cite{ellis} (see also Chavanis \cite{aaantonov}). Therefore, 
the maximization of $S$ at fixed $U$ and ${\cal M}$ guarantees (i) that 
the statistical equilibrium macrostate
is stable with respect to the perturbation on the microscopic scale
(Lynden-Bell thermodynamical stability) and (ii) that the
coarse-grained DF $\overline{f}$ is stable for the Vlasov equation
with respect to macroscopic perturbations (refined formal nonlinear
dynamical stability).  We again emphasize that it is only when the
initial DF takes two values $f_{0}$ and $0$ that the Lynden-Bell
entropy can be expressed in terms of the coarse-grained DF
$\overline{f}$, as in Eq. (\ref{eq:entropieVlasov}). In general, the
Lynden-Bell entropy is a functional of the probability distribution of
phase levels
\cite{super}. From Eq. (\ref{eq:entropieVlasov}), we write
the first order variations as
\begin{eqnarray}
\label{vr8}
\delta S-\beta\delta U-\alpha\delta {\cal M}  =0,
\end{eqnarray} 
where the inverse temperature $\beta=1/T$ and $\alpha$ are Lagrange
multipliers associated with the conservation of energy and
mass. Requiring that this entropy is stationary, one obtains the
following distribution~\cite{chavanis1,antoniazzi-06}
\begin{equation}
  \label{eq:barf}
  \bar{f}_{\text{QSS}}(\theta,p)=
  \frac{f_0}{1+e^{\beta (p^2/2 -M_x[\bar{f}_{\text{QSS}}]\cos\theta-M_y[\bar{f}_{\text{QSS}}]\sin\theta)+\alpha}}.
\end{equation}
As a general remark, it should be emphasized that
the above distribution differs from the Boltzmann-Gibbs one because of the
``fermionic'' denominator, which in turn arises because of the form
of the entropy. Morphologically, this distribution function is similar to
the Fermi-Dirac statistics so that several
analogies with the quantum mechanics setting are to be expected. Notice also that the magnetization is
related to the distribution function by Eq. (\ref{magnetization}) and the problem hence amounts to solving  
an {\it integro-differential} system. In doing so, we
have also to make sure that the critical point corresponds to an entropy maximum, not to a
minimum or a saddle point. 
Let us now insert expression (\ref{eq:barf}) into the energy and
normalization constraints and use the definition of
magnetization (\ref{magnetization}). Further, defining $\lambda=e^{\alpha}$ and
${\mathbf m}=(\cos \theta, \sin \theta)$ yields:   

\begin{eqnarray}
\label{eq:cond0}
&f_0& \sqrt{\frac{{2}}{{\beta}}}  \int {\mathrm d} \theta  I_{-1/2}\left(\lambda  e^{-\beta {\mathbf M} \cdot {\mathbf
m}}\right) = 1, \\
\label{eq:cond3}
&f_0& \frac{1}{2} \left (\frac{2}{\beta}\right )^{3/2} \int {\mathrm d} \theta
I_{1/2}\left(\lambda e^{-\beta {\mathbf M} \cdot {\mathbf m}}\right)
= U+\frac{M^2 - 1}{2}, \nonumber\\
\label{eq:cond1}
&f_0& \sqrt{\frac{{2}}{{\beta}}}   \int {\mathrm d} \theta \cos \theta
I_{-1/2}\left(\lambda e^{-\beta {\mathbf M} \cdot {\mathbf m}}\right)
= M_x, \nonumber \\
\label{eq:cond2}
&f_0& \sqrt{\frac{{2}}{{\beta}}} \int {\mathrm d} \theta \sin \theta
I_{-1/2}\left(\lambda e^{-\beta {\mathbf M} \cdot {\mathbf m}}\right)
= M_y, \nonumber
\end{eqnarray}
where we have defined the Fermi integrals 
\begin{eqnarray}
\label{es5}
I_{n}(t)=\int_{0}^{+\infty}\frac{x^{n}}{1+t e^{x}}dx.
\end{eqnarray}
We have the asymptotic behaviours for $t\rightarrow 0$:
\begin{eqnarray}
\label{deg4}
I_{n}(t)\sim\frac{(-\ln t)^{n+1}}{n+1}, \qquad (n>-1),
\end{eqnarray}
and for $t\rightarrow +\infty$
\begin{eqnarray}
\label{d4}
I_{n}(t)\sim\frac{\Gamma(n+1)}{t}, \qquad (n>-1).
\end{eqnarray}

The magnetization in the QSS, $\MQSS=M[\bar{f}_{\text{QSS}}]$, and the
values of the multipliers are hence obtained by numerically solving
the above coupled implicit equations. It should be stressed
that multiple local maxima of the entropy are in principle present
when solving the variational problem, thus resulting in a rich zoology
of phase transitions. This issue has been addressed in
\cite{chavanis1, antoniazzi-06} and more recently in
\cite{yama}, to which the following discussion refers to.

It is important to note that, in the two-levels approximation, the
Lynden-Bell equilibrium state depends only on two control parameters
$(U,f_0)$ \footnote{These parameters are related to those introduced
in \cite{chavanis1} by $U=\epsilon/4+1/2$, $\beta=2\eta$,
$f_0=\eta_{0}/N=\mu/(2\pi)$, $k=2\pi/N$, $x=\Delta\theta$,
$y=(2/\pi)\Delta p$ and the functions $F$ in \cite{antoniazzi-06} are
related to the Fermi integrals by
$F_{k}(1/y)=2^{(k+1)/2}yI_{(k-1)/2}(y)$.}. This is valid for {\it any}
initial condition with $f(\theta,p,t=0)\in \lbrace 0,f_0\rbrace$. This
general case will be studied in Section~\ref{diagr2} where we describe
the phase diagram in the $(f_0,U)$ plane. Now, many numerical simulations
of the $N$-body system or of the Vlasov equation have been performed
starting from a family of rectangular water--bag distributions. The
latter correspond to assuming a constant value $f_0$ inside the
phase-space domain $D$:
\begin{equation}
\label{water--bag_rect}
D = \{(\theta,p)\in [-\pi,\pi] \times [-\infty,\infty]~|~ |\theta|<\Delta\theta, ~|p|<\Delta p\},
\end{equation}
where $0\leq\Delta\theta\leq\pi$ and $\Delta p \geq 0$. The
normalization condition results in $f_0=1/(4\Delta\theta\Delta p)$.
Notice that, for this specific choice, the initial magnetization
$M_{0}$ and the energy density $U$ can be expressed as functions of
$\Delta\theta$ and $\Delta p$ as
\begin{equation}
\label{dtm}
M_{0} = \dfrac{\sin(\Delta\theta)}{\Delta\theta} ,
\quad
U = \dfrac{(\Delta p)^{2}}{6} + \dfrac{1-(M_{0})^{2}}{2}~.
\end{equation}
For the case under scrutiny, $0 \leq M_{0} \leq 1$ and $U \geq
U_{MIN}(M_0)\equiv (1 - M_0^2)/2$. The variables $(M_0,U)$ are
therefore used to specify the initial configuration and hereafter
assumed to define the relevant parameters space. This particular but
important case will be studied specifically in Section~\ref{diagr1} where we
illustrate the phase diagram in the $(M_0,U)$ plane for the rectangular
water--bag initial condition. Before that, we analytically study the
stability of the Lynden-Bell homogeneous phase: two important
limits, namely the non degenerate and the completely degenerate ones, are considered. 
We also discuss the condition for the existence of a
homogeneous, non--equilibrium phase.  

\section{Properties of the homogeneous Lynden-Bell distribution}
\label{homogeneous}

If we consider spatially homogeneous configurations ($M_{QSS}=0$), the
Lynden-Bell distribution becomes
\begin{equation}
  \label{vr9}
  \bar{f}_{\text{QSS}}(p)=
  \frac{f_0}{1+\lambda e^{\beta p^2/2}}.
\end{equation}
Using Eqs. (\ref{eq:cond0}), the relation between the inverse
temperature $\beta$ and the energy $U$ is given in parametric form by
\begin{eqnarray}
\label{cal} 
U - \frac{1}{2} =  \frac{1}{8\pi^{2}f_0^2}\frac{I_{1/2}(\lambda)}{I_{-1/2}(\lambda)^3},\qquad 
\beta = 8 \pi^2 f_0^2  I_{-1/2}(\lambda)^2.
\end{eqnarray}
This defines the series of equilibria $T(U)$ for fixed $f_0$
parametrized by $\lambda$ (see Fig. 3 in \cite{chavanis1}). The
stable part of the series of equilibria is the caloric curve.  Note
that the temperature $T$ is a Lagrange multiplier associated with the
conservation of energy in the variational problem (\ref{vr8}). It also
has the interpretation of a kinetic temperature in the Fermi-Dirac
distribution (\ref{vr9}). If we start from a water--bag initial
condition, recalling that $f_0=1/(4\Delta\theta\Delta p)$ and $(\Delta
p)^2 = 6\lbrack U-(1-M_0^2)/2\rbrack$, we can express $f_0$ as a
function of $M_0$ and $U$ by
\begin{eqnarray}
\label{f0mu}
f_0^2=\frac{1}{48\lbrack (2U-1)(\Delta\theta)^2+\sin^{2}\Delta\theta\rbrack},
\end{eqnarray}
where $\Delta\theta$ is related to $M_{0}$ by
Eq. (\ref{dtm}). Inserting this expression in Eqs. (\ref{cal}), we
obtain after some algebra the caloric curve $T(U)$ for fixed $M_0$
parametrized by $\lambda$:
\begin{eqnarray}
\label{calb}
&& U-\frac{1}{2}=\frac{\sin^{2}\Delta\theta}{\frac{\pi^{2}}{6}\frac{I_{-1/2}
(\lambda)^3}{I_{1/2}(\lambda)}-2(\Delta\theta)^2}, \nonumber \\
&& \beta=\frac{1}
{\sin^{2}\Delta\theta}\left (\frac{\pi^{2}}{6}I_{-1/2}
(\lambda)^2-2(\Delta\theta)^{2}\frac{I_{1/2}(\lambda)}{I_{-1/2}(\lambda)}\right ).
\end{eqnarray}
Eqs. (\ref{cal}) can be rewritten
\begin{equation}
\label{G1}
 (U-\frac{1}{2}) 8 \pi^2 f_0^2=G(\lambda) \equiv \frac{I_{1/2}(\lambda)}{I_{-1/2}(\lambda)^{3}},
\end{equation}
where $G(\lambda)$ is a universal function monotonically increasing
with $\lambda$ (see Fig. 2 of \cite{chavanis1}).  A solution of the above equation certainly exists
provided:
\begin{equation}
\label{G2}
(U-\frac{1}{2}) 8 \pi^2 f_0^2 \ge G(0).
\end{equation}
To compute $G(0)$ we use the asymptotic expansions (\ref{deg4}) and
(\ref{d4}) of the Fermi integrals. This yields $G(0)=1/12$. Therefore, the homogeneous Lynden-Bell 
distribution with fixed $f_0$ exists only for \cite{chavanis1}: 
\begin{eqnarray}
\label{miny}
U\ge U_{min}(f_0)\equiv\frac{1}{96\pi^2 f_0^{2}}+\frac{1}{2}.
\end{eqnarray}
For the rectangular water--bag initial condition, using Eqs. (\ref{f0mu}) and
(\ref{miny}), we here find that the homogeneous Lynden-Bell distribution with fixed $M_0$ exists only for: 
\begin{eqnarray}
\label{cn5}
U\ge U_{min}(M_0)\equiv \frac{1}{2}\left (\frac{\sin^{2}\Delta\theta}{\pi^{2}-(\Delta\theta)^{2}}+1\right ).
\end{eqnarray}
This result can also be obtained
from Eq. (\ref{calb}) by taking the limit $\lambda\rightarrow 0$.

\begin{figure}[htbp]
\centering
\vspace*{3em}
\includegraphics[width=10cm]{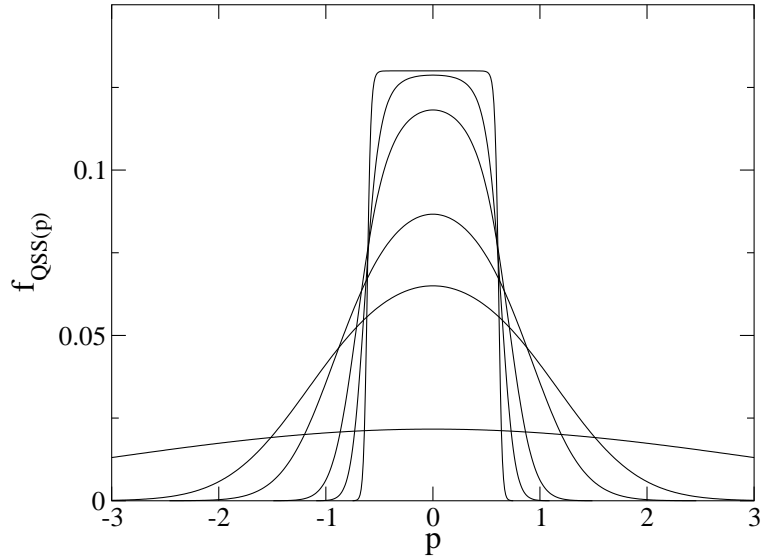}
\caption{Spatially homogeneous Lynden-Bell distribution function for increasing values of 
$\lambda$ (top to bottom). For $\lambda=0$, the distribution reduces to a step function 
(completely degenerate) and for $\lambda\rightarrow +\infty$, it becomes equivalent 
to the Maxwell-Boltzmann distribution (non degenerate). In the figure, we have taken 
$f_0=0.13$ and $\beta$ has been calculated from Eq. (\ref{cal}).}
\label{fqss}
\end{figure}

Let us now describe more precisely the asymptotic limits of the Fermi-Dirac distribution 
(see Fig. \ref{fqss}):

{\it Non degenerate limit:} In the limit $\lambda\rightarrow +\infty$, the Lynden-Bell 
distribution reduces to the Maxwell-Boltzmann distribution  
\begin{eqnarray}
\label{d1}
\overline{f}_{QSS}(p)=\left (\frac{\beta}{2\pi}\right )^{1/2}e^{-\beta {p^{2}}/{2}}.
\end{eqnarray}
Since $\overline{f}\ll f_{0}$, this corresponds to a dilute limit (or
to a non degenerate limit if we use the terminology of quantum
mechanics). The non degenerate limit corresponds, for a given value of
$f_0$, to $\lambda\rightarrow +\infty$, $\beta\rightarrow 0$ and
$U\rightarrow +\infty$. For $f_0\rightarrow +\infty$, we are always in
the non degenerate limit, for any $\beta$ and $U$. In that case, the
caloric curve takes the ``classical'' expression
\begin{eqnarray}
\label{cc10}
U=\frac{1}{2\beta}+\frac{1}{2},
\end{eqnarray}
a relation that can be directly obtained from Eq. (\ref{d1}).  For the
water--bag initial condition, the non degenerate limit
$\lambda\rightarrow +\infty$ corresponds, for a given $M_0$, to
$U\rightarrow +\infty$. The non degenerate limit $f_0\rightarrow
+\infty$ corresponds to $\Delta\theta=0$ leading to $M_0=1$ for any
$U$ \footnote{It also corresponds to $\Delta p=0$ leading to
$U=U_{MIN}(M_0)$ for any $M_0$. However, in that case the homogeneous
phase does not exist since $U_{MIN}(M_0)\le U_{min}(M_0)$ so this case
will not be considered here.}. In the non degenerate limit, the
Lynden-Bell statistical equilibrium state describing the QSS has the
{\it same} structure as the Boltzmann distribution describing the
collisional statistical equilibrium state (but with,
of course, a completely different interpretation).

{\it Completely degenerate limit:} In the limit $\lambda\rightarrow 0$, the Lynden-Bell
distribution (\ref{vr9}) reduces to the Heaviside function
\begin{eqnarray}
\overline{f}_{QSS}(p)=\Biggl\lbrace \begin{array}{cc}
f_{0}   & \qquad (p< p_{F}), \\
0  & \qquad (p\ge p_{F}),
\end{array}
\label{deg1}
\end{eqnarray}
where 
\begin{eqnarray}
\label{deg2}
p_{F}=\frac{1}{4\pi f_0},
\end{eqnarray}
is a maximum velocity.  The distribution (\ref{deg1}) is similar to
the Fermi distribution in quantum mechanics and $p_{F}$ is similar to
the Fermi velocity. Thus, the limit $\lambda\rightarrow 0$ corresponds
to a completely degenerate limit in the quantum mechanics
terminology. The completely degenerate limit corresponds to
$\lambda\rightarrow 0$, $\beta\rightarrow +\infty$ and $U\rightarrow
U_{min}(f_0)$ given by (\ref{miny}).  This result can be directly
obtained from Eq. (\ref{deg1}). This is the minimum energy of the
homogeneous Lynden-Bell distribution for a fixed $f_0$. This is also
the minimum energy of any {\it homogeneous} distribution with $f\in
\lbrace 0,f_{0}\rbrace$. It corresponds to a water--bag initial
condition with zero magnetization $M_{0}=0$. If we start from a water
bag initial condition, the completely degenerate limit corresponds to
$M_{0}=0$ for any $U$ and to $U=U_{min}(M_0)$ for any $M_0$. A {\it
stable} water--bag initial condition with $M_{0}=0$ is a maximum
Lynden-Bell entropy state, so it does not mix at all. 

In conclusion, we have in the general case, using the $(U,f_0)$ variables:

\begin{itemize}
\item non degenerate limit
      \begin{itemize}
      \item $U \to +\infty$ for any $f_0$
      \item $f_0 \to +\infty$ for any $U$
      \end{itemize}
\item completely degenerate limit
      \begin{itemize}
      \item $U = U_{min} (f_0)$ for any $f_0$
      \end{itemize}
\end{itemize}

For the water--bag initial condition, using the $(U,M_0)$ variables, we have:

\begin{itemize}
\item non degenerate limit
      \begin{itemize}
      \item $U \to +\infty$ for any $M_0$
      \item $M_0=1$ for any $U$
      \end{itemize}
\item completely degenerate limit
      \begin{itemize}
      \item $M_0=0$ for any $U$
      \item $U=U_{min}(M_0)$ for any $M_0$
      \end{itemize}
\end{itemize}

\section{Stability of the Lynden-Bell homogeneous phase}
\label{stability}

We have seen that the maximization problem (\ref{vr7}) provides a
condition of thermodynamical stability (in Lynden-Bell's sense) and a
condition of nonlinear dynamical stability with respect to the Vlasov
equation. We thus have to select the {\it maximum} of $S$ at fixed
$U$, ${\cal M}$. Indeed, a saddle point of $S$ is unstable and cannot be
obtained as a result of a violent relaxation. Let us consider the 
minimization problem
\begin{eqnarray}
\label{stab1}
\min_{\overline{f}}\lbrace F[\overline{f}]= U
[\overline{f}]-T S[\overline{f}]\quad |\quad  {\cal M}[\overline{f}]=1\rbrace.
\end{eqnarray} 
The criterion (\ref{vr7}) can be viewed as a criterion of
microcanonical stability and the criterion (\ref{stab1}) as a
criterion of canonical stability where $F$ is interpreted as a free
energy. Quite generally, the solutions of (\ref{stab1}) are solutions
of (\ref{vr7}) but the reciprocal is wrong in case of ensemble
inequivalence. Therefore, in the general case,
criterion (\ref{stab1}) forms a {\it sufficient} (but not necessary)
condition of thermodynamical and formal nonlinear dynamical
stability. In the present case, it can be shown that,
if we restrict ourselves to spatially homogeneous
solutions \footnote{Concerning spatially inhomogeneous
solutions, the microcanonical and canonical ensembles are not
equivalent in the region of first order phase transition. This
important point will be further developed in a future
contribution.}, the ensembles are equivalent so that the set of
solutions of (\ref{stab1}) {\it coincides} with the set of solutions
of (\ref{vr7}). Therefore, considering homogeneous
states, criterion (\ref{stab1}) forms a necessary and sufficient
condition of thermodynamical and formal nonlinear dynamical
stability. We shall therefore consider in the following the
minimization problem (\ref{stab1}), which has been studied in \cite{barre-06,chavanis} 
for general functionals of the form $S[f]=-\int C(f)d\theta dp$ where $C$ is a convex function. 
A simple stability criterion has been obtained in the case where
the steady state is spatially homogeneous, which can
be expressed in terms of the distribution function as
\cite{barre-06}:
\begin{eqnarray}
\label{cd1}
1+{\pi}\int_{-\infty}^{+\infty}\frac{f'(p)}{p}dp\ge 0.
\end{eqnarray}
It was shown in \cite{chavanis} that the same criterion can be expressed simply 
in terms of the  density as
\begin{eqnarray}
\label{vs1}
c_{s}^{2}\ge \frac{1}{2},
\end{eqnarray}
where $c_{s}^{2}=p'(\rho)$ is the velocity of sound in the
corresponding barotropic gas. The equivalence between the criteria
(\ref{cd1}) and (\ref{vs1}) is proven in \cite{chavanis}. In this
Section, we apply these criteria to the Lynden-Bell distribution
(\ref{vr9}). It is shown in \cite{chavanis1} that the criteria (\ref{cd1})
and (\ref{vs1}) can be rewritten:
\begin{eqnarray}
\label{vs7}
I_{-1/2}(\lambda)\lambda |I_{-1/2}'(\lambda)|\le \frac{1}{(2\pi f_0)^2}.
\end{eqnarray}
If the DF satisfies (\ref{vs7}) $\Leftrightarrow $ (\ref{cd1})
$\Leftrightarrow $ (\ref{vs1}),  then it is (i) Lynden-Bell
thermodynamically stable (ii) formally nonlinearly dynamically
stable. Otherwise, it is (i) Lynden-Bell thermodynamically unstable
(ii) linearly dynamically unstable \cite{chavanis}. For given $f_0$, the
relation (\ref{vs7}) determines the range of $\lambda$ for which the
homogeneous distribution is stable/unstable. Then, using
Eqs. (\ref{cal}), we can determine the range of
corresponding energies. Specifically, the critical curve
$U_{c}(f_0)$ separating stable and unstable homogeneous Lynden-Bell distributions 
is given by the parametric equations \cite{chavanis1}:
\begin{eqnarray}
\label{vs7b}
I_{-1/2}(\lambda)\lambda |I_{-1/2}'(\lambda)|= \frac{1}{(2\pi f_0)^2}, 
\qquad U-\frac{1}{2}= \frac{1}{8 \pi^2 f_0^2} \frac{I_{1/2}(\lambda)}{I_{-1/2}(\lambda)^{3}}
\end{eqnarray}
where $\lambda$ goes from $0$ (completely degenerate) to $+\infty$
(non degenerate).  This leads to the phase diagram reported in
Fig. \ref{muepsilon}. In fact, the criteria (\ref{cd1}) and
(\ref{vs1}) only prove that $f$ is a {\it local} entropy maximum at
fixed mass and energy. If several local entropy maxima are found, we
must compare their entropies to determine the stable state (global
entropy maximum) and the metastable states (secondary entropy
maxima). For systems with long-range interactions, metastable states
have in general very long lifetimes, scaling like $e^{N}$, so that
they are stable in practice and must absolutely be
taken into account \cite{lifetimeantoni,lifetime}.

For $f_0\rightarrow +\infty$, we are in the non degenerate limit
$\lambda\rightarrow +\infty$ and the stability criterion (\ref{vs7}) for the
homogeneous phase becomes
\begin{eqnarray}
\label{lc1}
U\ge U_{c}= \frac{3}{4}.
\end{eqnarray}
This returns the well-known nonlinear dynamical stability criterion
(with respect to the Vlasov equation) of a homogeneous system with
Maxwellian distribution function (see, e.g.,
\cite{barre-06,chavanis}).  This also coincides with the ordinary thermodynamical
stability criterion applying to the {\it collisional} regime, for
$t\rightarrow +\infty$, where the statistical equilibrium state is the
Boltzmann distribution for $f$ (without the bar!).

On the curve $U=U_{min}(f_0)$, we are in the completely degenerate limit 
$\lambda\rightarrow 0$ and the stability criterion (\ref{vs7}) for the
homogeneous phase becomes
\begin{eqnarray}
\label{lc3}
f_0\le (f_{0})_c=\frac{1}{2\pi\sqrt{2}}, \quad {i.e}\quad U\ge U_{c}=\frac{7}{12}.
\end{eqnarray}
This is the well-known nonlinear dynamical stability criterion
(with respect to the Vlasov equation) of the water--bag distribution
(see, e.g., \cite{barre-06,chavanis}).

If we start from a rectangular water--bag initial condition and use the $(U,M_0)$
variables, we must express $f_0$ in terms of $U$ and $M_0$ using
Eq. (\ref{f0mu}). Then, the critical curve $U_{c}(M_0)$ separating
stable and unstable homogeneous Lynden-Bell distributions is given by
the parametric equations
\begin{eqnarray}
\label{vs7bb}
I_{-1/2}(\lambda)\lambda |I_{-1/2}'(\lambda)|= \frac{1}{(2\pi f_0)^2}, \qquad U-\frac{1}{2}= 
\frac{1}{8 \pi^2 f_0^2} \frac{I_{1/2}(\lambda)}{I_{-1/2}(\lambda)^{3}},
\end{eqnarray}
\begin{eqnarray}
 \label{f0mub}
f_0^2=\frac{1}{48\lbrack (2U-1)(\Delta\theta)^2+\sin^{2}\Delta\theta\rbrack}, 
\qquad M_{0} = \dfrac{\sin(\Delta\theta)}{\Delta\theta} ,
\end{eqnarray}
where $\lambda$ goes from $0$ (completely degenerate) to $+\infty$
(non degenerate).  This leads to the phase diagram reported in
Fig. \ref{fig:phase-diagram}.  For $M_0=1$, we get $\lambda \rightarrow
+\infty$ so we are in the non degenerate limit and the critical energy
is $U_c=3/4$. For $M_0=0$, we get $\lambda=0$ so we are in the
completely degenerate limit and the critical energy is $U_c=7/12$.

\section{The rectangular water--bag initial condition: phase diagram in the $(M_0,U)$ plane}
\label{diagr1}

\begin{figure}[htbp]
  \centering \vspace*{3em} \includegraphics[width=10cm]{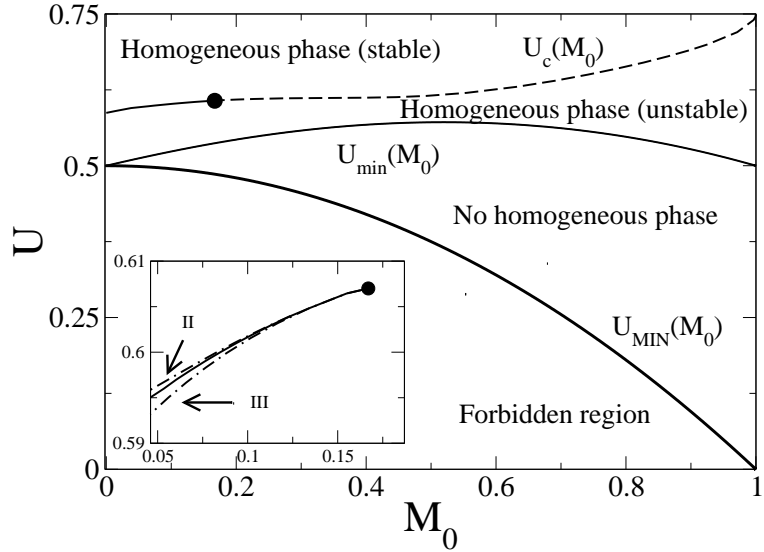}
  \caption{Theoretical phase diagram in the control parameter plane
  $(M_{0},U)$ for a rectangular water--bag initial profile. The dashed
  line $U_c(M_0)$ stands for the second order phase transition, while its
  continuation as a full line refers to the first order phase transition. 
  The full dot is the tricritical point. The region below the thick full line 
  $U_{MIN}(M_0)$ corresponds to the forbidden domain of the parameter space. 
  The region of existence of the homogeneous solution is delimited from below 
  by the thin full line  $U_{min}(M_0)$ (see Eq. (\ref{cn5})). Inset:
  zoom of the first order transition region. Dash-dotted lines
  represent the limits of the metastability region. In region (II), delimited from
  above by the upper dash-dotted line and from below by the full line,  the  
  homogeneous phase is fully stable and the inhomogeneous phase is metastable. 
  In region (III), located below the full line and above the lower dash-dotted line, 
  the  homogeneous phase is metastable and the inhomogeneous phase is fully stable. 
  These labels will appear again in Fig. \ref{meta}, in connection
  with the discussion of the general case.
  } 
  \label{fig:phase-diagram}
\end{figure}

We first comment on the structure of the phase diagram in the
$(M_0,U)$ plane when we start from a water--bag initial condition.  In
Fig. \ref{fig:phase-diagram} the transition line $U_{c}(M_0)$ divides
the region of the plane where a homogeneous ($\MQSS=0$) state is
predicted to occur (upper area), from that (lower area) associated to a non-homogeneous phase
($\MQSS \ne 0$).  Along the transition line two distinct segments can
be isolated: the dashed line corresponds to a second order phase
transition, the full line refers to a first order phase transition.
First and second transition lines merge together at a tricritical
point, approximately located at $(M_{0},U)=(0.17,0.61)$. The lateral edges of the metastability 
region associated to the first order transition line are also reported in the inset of 
Fig. \ref{fig:phase-diagram}.

The correctness of the above analysis is assessed in \cite{yama} where
numerical simulations of the HMF model (\ref{eq:ham}) are performed
for different values of the system size $N$. The transition predicted
in the realm of Lynden Bell's theory are indeed numerically observed,
thus confirming the adequacy of the proposed interpretative
scenario. Moreover, the coexistence of homogeneous and inhomogeneous
phases, corresponding to different local maxima of the entropy, is
also reported in \cite{yama} where the probability distribution
function of $M$ is reconstructed.

\begin{figure}[htbp]
  \centering \vspace*{2em} \includegraphics[width=10cm]{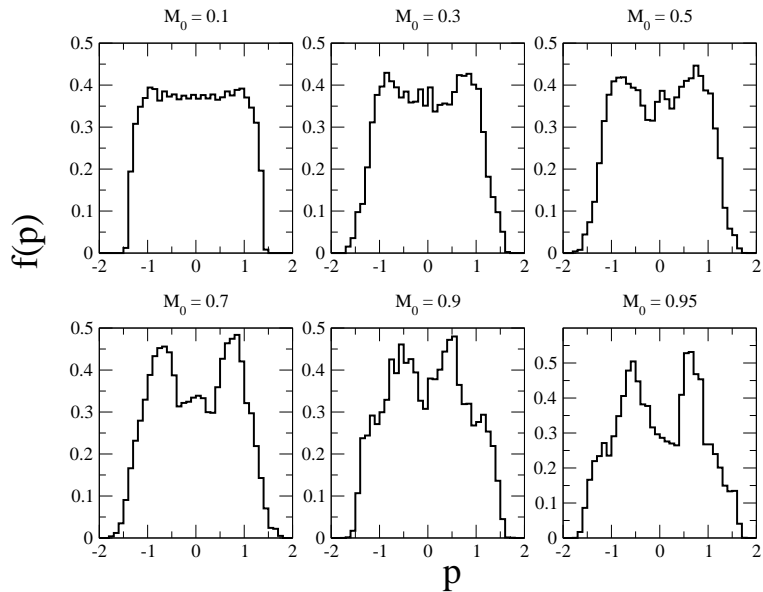}
  \caption{Probability distribution function of velocities $f(p)$, for
  U=0.85 and different initial magnetization, as reported in the
  legend of each panel. } 
  \label{fig:simulations}
\end{figure}

For all values of the energy larger than $U_{c}(M_0)$ (where the
homogeneous phase is stable), the systems can potentially encounter a
homogeneous quasi-stationary phase which slows down the approach to
the thermodynamic equilibrium. Notice that above $U_c=0.75$, the
equilibrium value of the magnetization is also zero: there is hence no
macroscopic transition for $M(t)$ of the type displayed in
Fig. \ref{fig:QSS}. One has therefore to rely on other, quantitative,
indicators to monitor the dynamical state of the system
\cite{morelli}, and eventually assess the presence of a QSS. This
explains why in the past QSSs regimes where believed to be localized
only in specific energy windows below the critical threshold
$U_c$. Significant deviations from equilibrium are instead detected
even for $U>U_c$, as reported in Fig.~\ref{fig:simulations}. Single
particle velocity distributions reconstructed from direct $N$--body
simulations at $U=0.85$ display a bumpy profile, analogous to the one
discussed in \cite{antoniazzi-06} for the reference energy
$U=0.69<U_c$. Interestingly, for specific choices of the initial
magnetization, the two bumps are even more pronounced than those
analyzed in \cite{antoniazzi-06}.  The presence of these bumps shows
that relaxation is incomplete. These bumps correspond to the
``vortices'' (or phase space clumps) visible in
Fig. \ref{fig:rectangular}. This state is stationary for the
Vlasov equation, but Quasi-Stationary for the $N$-body simulation. Hence,
in the long run, the two ``vortices'' will merge, due to finite $N$ effects, 
as the HMF system proceeds towards Boltzmann-Gibbs equilibrium.

\section{The general case: phase diagram in the $(f_0,U)$ plane}
\label{diagr2}

In the two-levels approximation, the Lynden-Bell equilibrium state
depends only on two control parameters $(U,f_0)$. This is valid for
{\it any} initial condition with $f(\theta,p,t=0)\in \lbrace
0,f_0\rbrace$, whatever the number of patches and
their shape. The variables $(U,M_{0})$ used in the
previous section are valid only for a rectangular water--bag
configuration. Furthermore, we note that two configurations with
different values of $(U,M_{0})$ in a rectangular water--bag
configuration can correspond to the {\it same} values of $(U,f_0)$,
hence to the {\it same} Lynden-Bell equilibrium state. To avoid any
redundance, it is preferable to use the general variables $(U,f_0)$. 
Therefore, in the following, we shall discuss the general phase
diagram in the $(U,f_0)$ plane \cite{chavanis1} and compare it with
the one in the $(U,M_{0})$ plane for the rectangular water--bag
assumption \cite{yama}.

\begin{figure}[htbp]
\centering
\vspace*{3em}
\includegraphics[width=10cm]{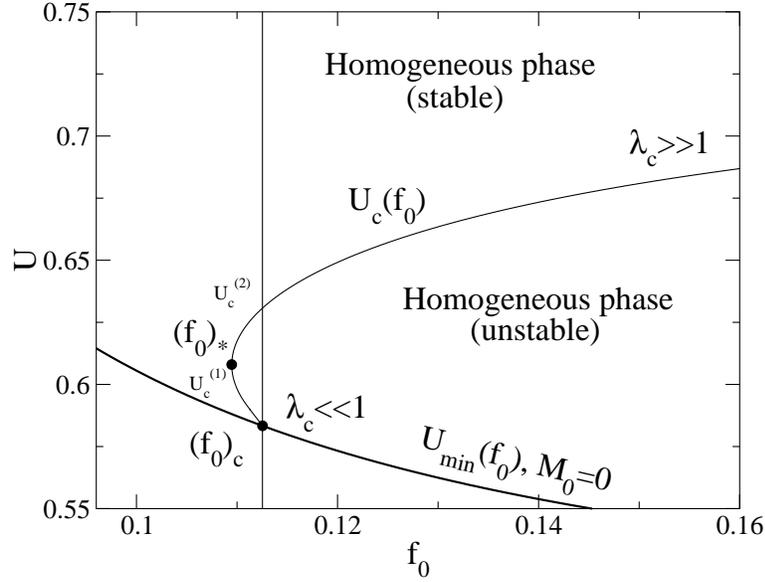}
\caption{Stability diagram of the homogeneous phase in the $(f_0,U)$ plane. The 
homogeneous phase only exists above the line $U_{min}(f_0)$. It is stable above the line 
$U_{c}(f_0)$ and unstable below it. For fixed $f_{0}\in \lbrack (f_{0})_{*},(f_{0})_{c}\rbrack$, 
there is a re-entrant phase as we progressively lower the energy: the homogeneous phase is stable 
for $U>U_{c}^{(2)}(f_0)$, unstable for
$U_{c}^{(2)}(f_0)>U>U_{c}^{(1)}(f_0)$ and {\it stable again} for
$U_{c}^{(1)}(f_0)>U\ge U_{min}(f_0)$. }
\label{muepsilon}
\end{figure}

The stability diagram of the homogeneous Lynden-Bell distribution
(\ref{vr9}) is plotted in Fig.  \ref{muepsilon} in the $(f_0,U)$
plane.  The representative curve $U_{c}(f_0)$ marks the separation
between the stable (maximum entropy states) and the unstable (saddle
points of entropy) regions.  We have also plotted the minimum
accessible energy for the homogeneous phase $U_{min}(f_0)$.  Here, the
term ``unstable'' means that the homogeneous Lynden-Bell distribution
is not a maximum entropy state, i.e. (i) it is not the most mixed
state (ii) it is dynamically unstable with respect to the Vlasov
equation. Hence, it should not be reached as a result of violent
relaxation. One possibility is that the system converges to the
spatially {\it inhomogeneous} Lynden-Bell distribution (\ref{vr9})
with $M_{QSS}\neq 0$ which is the maximum entropy state (most mixed)
in that case. Another possibility, always to consider, is that the
system does not converge towards the maximum entropy state, i.e. the
relaxation is {\it incomplete} \cite{next05}.

Let us enumerate some properties of the $(f_0,U)$ phase diagram of Fig.~\ref{muepsilon}. 
For $U>U_c=3/4$ (supercritical energies), the homogeneous phase is
always stable (maximum entropy state) for any $f_0$
(recall that the curve $U_{c}(f_{0})\rightarrow 3/4$
for $f_0\rightarrow +\infty$). On the other hand, there exists a
critical point (we shall see later that it corresponds
to the tricritical point of Fig. \ref{fig:phase-diagram}) located at
\begin{eqnarray}
\label{vs8}
(f_{0})_{*}=0.10947..., \qquad U_{*}=0.608...
\end{eqnarray}
For $f_0<(f_0)_{*}$, the homogeneous phase is always stable (maximum
entropy state) for any $U\ge U_{min}(f_0)$. For $f_0>(f_0)_{c}$, the
homogeneous phase is stable for $U>U_{c}(f_0)$ and unstable for
$U_{min}(f_0)\le U<U_{c}(f_0)$. This range of parameters corresponds
to a second order phase transition. For $(f_0)_{*}<f_0<(f_0)_{c}$,
there is an interesting regime with a ``re-entrant'' phase
\cite{reentrant} . The homogeneous phase is
stable for $U>U_{c}^{(2)}(f_0)$, unstable for
$U_{c}^{(2)}(f_0)>U>U_{c}^{(1)}(f_0)$ and {\it stable again} for
$U_{c}^{(1)}(f_0)>U\ge U_{min}(f_0)$. This range of parameters
corresponds to a first order phase transition.

\begin{figure}[htbp]
\centering
\vspace*{3em}
\includegraphics[width=10cm]{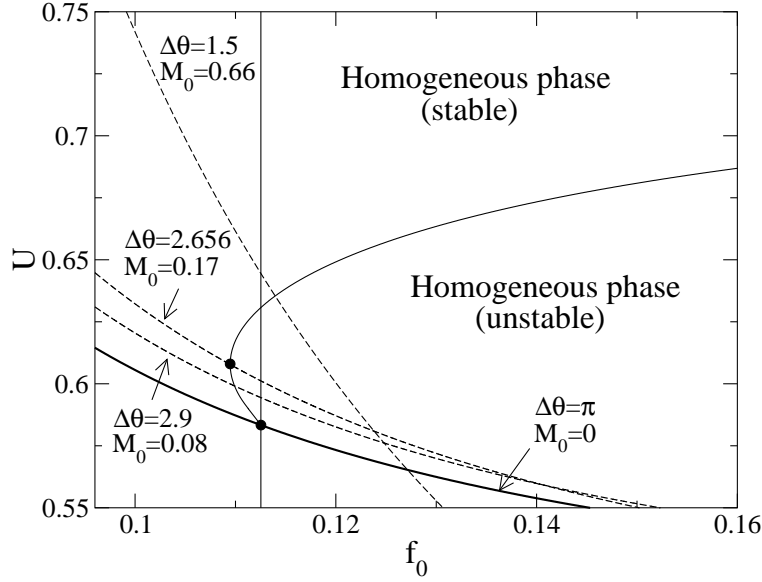}
\caption{Iso-$M_0$ lines in the $(f_0,U)$ phase diagram. 
This graphical construction allows one to make the connection between the $(f_0,U)$  
phase diagram of Fig.~\ref{muepsilon} and the $(M_0,U)$ phase diagram of Fig.~\ref{fig:phase-diagram}. 
We can vary the energy at fixed initial magnetization by following a dashed line. The 
intersection between the dashed line and the curve $U_{min}(f_{0})$ determines the minimum energy 
$U_{min}(M_{0})$ of the homogeneous phase. The intersection between the dashed line and the curve 
$U_{c}(f_{0})$ determines the energy $U_{c}(M_{0})$ below which the homogeneous phase becomes unstable.}
\label{muepsilonINTER}
\end{figure}

\begin{figure}[htbp]
\centering
\vspace*{3em}
\includegraphics[width=10cm]{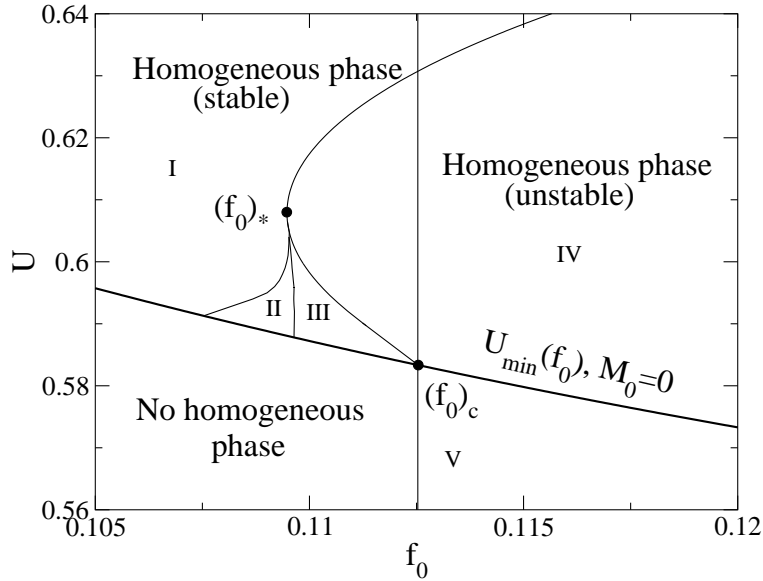}
\caption{Zones of metastability in the $(f_0,U)$ phase diagram. In region (I), the homogeneous 
phase is fully stable and the inhomogeneous phase is inexistent. In region (II)  the  
homogeneous phase is fully stable and the inhomogeneous phase is metastable. In region (III)  
the  homogeneous phase is metastable and the inhomogeneous phase is fully stable. 
In region (IV) the homogeneous phase is unstable and the inhomogeneous phase is fully stable. 
The three curves separating these regions connect themselves at the tricritical point.
In region (V) the homogeneous phase is inexistent.  The corresponding
caloric curves as well as the absolute minimum energy $U_{MIN}(f_0)$
of the inhomogeneous phase will be determined in a future contribution.}
\label{meta}
\end{figure}

To make the connection between the phase diagram $(f_0,U)$ obtained in
\cite{chavanis1} and the phase diagram $(M_0,U)$ obtained in
\cite{yama}, we can plot the iso-$M_{0}$ lines in the
$(f_0,U)$ phase diagram. If we fix the initial magnetization
$M_{0}$, or equivalently if we fix the parameter $\Delta\theta$, the
relation between the energy $U$ and
$f_0$ is
\begin{eqnarray}
\label{cn1}
U_{\Delta\theta}(f_0)=\frac{1}{6(4\Delta\theta f_0)^{2}}-\frac{1}{2} 
\left (\frac{\sin \Delta\theta}{\Delta\theta}\right )^{2}+\frac{1}{2}.
\end{eqnarray}
Therefore, the iso-$M_{0}$ lines are of the form
\begin{eqnarray}
\label{cn2}
U_{\Delta\theta}(f_0)=\frac{A(\Delta\theta)}{f_0^{2}}-B(\Delta\theta),
\end{eqnarray}
with $A(\Delta\theta)=\frac{1}{6(4\Delta\theta)^2}$ and
$B(\Delta\theta)=\frac{1}{2} \left (\frac{\sin
\Delta\theta}{\Delta\theta}\right )^{2}-\frac{1}{2}$, which are easily 
represented in the $(f_0,U)$ phase diagram (see
Fig. \ref{muepsilonINTER}). As an immediate consequence of this
geometrical construction, we can recover the minimum energy of the
homogeneous phase for a fixed initial magnetization $M_{0}$ (or
$\Delta\theta$). Indeed, for a given $\Delta\theta$, the homogeneous
phase exists iff $U_{\Delta\theta}(f_0)\ge U_{min}(f_0)$ leading to
\begin{eqnarray}
\label{cn3}
f_0^{2}\le (f_0)_{\Delta\theta}^{2}\equiv \frac{\pi^{2}-(\Delta\theta)^{2}}{48\pi^2\sin^{2}\Delta\theta}.
\end{eqnarray}
This corresponds to $U\ge U_{min}(M_0)=U_{\Delta\theta}((f_0)_{\Delta\theta})$ leading to 
\begin{eqnarray}
\label{cn5_1}
U\ge U_{min}(M_0)=\frac{1}{2}\left (\frac{\sin^{2}\Delta\theta}{\pi^{2}-(\Delta\theta)^{2}}+1\right ),
\end{eqnarray}
which is identical to (\ref{cn5}). Note again that $\Delta\theta$ is
related to $M_{0}$ by Eq. (\ref{dtm}). Figure \ref{muepsilonINTER} is
in good agreement with the structure of the phase diagram in the
$(U,M_0)$ plane. Indeed, along an iso-$M_0$ line, we find that for
large energies $U>U_{c}(M_0)$ the homogeneous phase is stable and for
low energies $U<U_{c}(M_0)$ the homogeneous phase becomes unstable. In
that case, there is no re-entrant phase.

In \cite{chavanis1}, only the stability of the homogeneous phase has
been studied, i.e. whether it is an entropy maximum at fixed mass and
energy or not. The question of its metastability, i.e. whether it is a
local entropy maximum with respect to the inhomogeneous phase, has not
been considered. However, considering Fig. \ref{muepsilonINTER} and
comparing with the results of
\cite{yama}, we conclude that there must exist zones of metastability
in the $(f_0,U)$ phase diagram. They have been
represented in Fig. \ref{meta}. In region (I) the inhomogeneous phase
does not exist while the homogeneous phase is fully stable. In region
(II), the inhomogeneous phase appears but is metastable while the
homogeneous phase is fully stable. In region (III), the inhomogeneous
phase becomes fully stable while the homogeneous phase becomes
metastable. In region (IV), the homogeneous phase becomes unstable
while the inhomogeneous phase is fully stable. The three curves
separating these regions connect themselves at a tricritical
point. This is clearly the same as in Fig. \ref{fig:phase-diagram}. 
Using Eqs. (\ref{vs8}), (\ref{cn1}), (\ref{dtm}) we find that it
corresponds to
\begin{eqnarray}
\label{cn6}
U_{*}=0.608...,  \qquad (M_{0})_{*}=0.1757...
\end{eqnarray}
with $\Delta\theta_{*}=2.656...$. Therefore, the phase
diagrams in $(f_{0},U)$ and $(M_{0},U)$ planes are fully
consistent. Note, however, that the physics is different whether we
vary the energy at fixed $f_0$ or at fixed $M_{0}$. In particular,
there is no ``re-entrant'' phase when we vary the energy at fixed
$M_{0}$ \cite{yama} while a ``re-entrant'' phase appears when we vary
the energy at fixed $f_{0}$ \cite{chavanis1}.

\section{Conclusions}
\label{concl}

In this paper, we have discussed the emergence of out-of-equilibrium
Quasi Stationary States (QSSs) in the Hamiltonian Mean Field (HMF)
model, a paradigmatic representative of systems with long-range
interactions. The analysis refers to a special class of initial
conditions in which particles are uniformly occupying a finite portion
of phase space and the distribution function takes only two values,
respectively $0$ and $f_0$. The energy can be independently fixed to
the value $U$.

The Lynden-Bell maximum entropy principle is here reviewed and shown
to result in a rich out--of--equilibrium phase diagram, which is
conveniently depicted in the reference plane $(f_0,U)$
\cite{chavanis1}. When considering a rectangular water--bag
distribution the concept of initial magnetization, $M_0$, naturally
arises as a control parameter and the different QSSs phases can be
represented in the alternative space $(M_0, U)$ \cite{yama}.  In both
settings first and second order phase transitions are found, which
merge together in a tricritical point. These findings have been tested
versus numerical simulation in \cite{yama}, where the adequacy of
Lynden--Bell theory was confirmed.

A formal correspondence between the two above scenarios is here drawn
and their equivalence discussed. It is worth mentioning that swapping
from one parametric representation to the other allows us to put the
focus on intriguingly different physical mechanisms, as it is the case
of the ``re-entrant'' phases discussed in Section \ref{diagr2}.

Further, in this paper, we have provided an analytical characterization of
the domain of existence of the Lynden-Bell spatially homogenous phase
and investigated its stability. Homogeneous QSS are also expected to
occur for $U>U_c=3/4$, a claim here supported by dedicated numerical
simulations.
   
Despite the fact that Lynden-Bell's theory results in an accurate tool
to explain the peculiar traits of QSSs in HMF dynamics, one should be
aware of the limitations which are intrinsic to this approach. Most
importantly, Lynden-Bell's recipe assumes that the system mixes well so
that the hypothesis of {\it ergodicity}, which motivates the
statistical theory (maximization of the entropy),
applies. Unfortunately, this is not true in general. Several example
of {\it incomplete} violent relaxation have been identified in stellar
dynamics and 2D turbulence (see some references in
\cite{next05}) for which the QSSs cannot be exactly described in term of a
Lynden-Bell distribution. Also in this case, however, the QSSs are
stable stationary solution of the Vlasov equation and novel analytical
strategies are to be eventually devised which make contact with the
underlying Vlasov framework.

\vspace*{0.5em}
\noindent
{\bf Acknowledgements}: 

D.F. and S.R. wish to thank A. Antoniazzi, F. Califano and Y. Yamaguchi for useful
discussions and long-lasting collaboration. This work is funded by the PRIN05 grant 
{\it Dynamics and thermodynamics of systems with long-range interactions}.


\begin{thebibliography}{99}

\bibitem{astro}  P.J. Peebles, \emph{The Large Scale Structure of the Universe}, 
(Princeton University Press, Princeton, NJ, 1980). 
\bibitem{turchetti} C. Benedetti, S. Rambaldi, G. Turchetti, Physica A {\bf 364}, 197 (2006); 
P.H. Chavanis, Eur. Phys. J. B {\bf 52}, 61 (2006).
\bibitem{barre} J. Barr{\'e}, T. Dauxois, G. de Ninno, D. Fanelli, S. Ruffo, Phys. Rev E {\bf 69}, 045501(R) (2004). 
\bibitem{Houches02} T. Dauxois et al., \emph{Dynamics and Thermodynamics 
of Systems with Long Range Interactions}, Lect. Not. Phys. {\bf 602}, Springer  (2002). 
\bibitem{debate} A. Rapisarda, A. Pluchino, Europhysics News {\bf 36}, 202 (2005); F. Bouchet, 
T. Dauxois and S. Ruffo, Europhysics News {\bf 37}, 9 (2006).
\bibitem{antoni-95} M. Antoni, S. Ruffo, Phys. Rev. E \textbf{52}, 2361 (1995).
\bibitem{Konishi} T. Tsuchiya, T. Konishi, N. Gouda, Phys. Rev. E {\bf  50}, 2607 (1994). 
\bibitem{ElskensEscande} Y.~Elskens, D.F.~Escande, {\em Microscopic
Dynamics of Plasmas and Chaos}, IoP Publishing, Bristol (2003).
\bibitem{chavanis} P.H. Chavanis, J. Vatteville, F. Bouchet, Eur. Phys. J. B 
{\bf 46}, 61 (2005) and references therein.
\bibitem{chavanis1}  P.H. Chavanis, Eur. Phys. J. B {\bf 53}, 487 (2006).
\bibitem{antoniazzi-06} A. Antoniazzi, D. Fanelli, J. Barr\'e, P.H. Chavanis, T. Dauxois, S. Ruffo, Phys. Rev. E
{\bf 75}, 011112 (2007).
\bibitem{yama} A. Antoniazzi, D. Fanelli, S. Ruffo, Y. Y. Yamaguchi, Phys. Rev. Lett. {\bf 99} 040601 (2007).
\bibitem{Latora} V. Latora, A. Rapisarda, S. Ruffo, Phys. Rev. Lett. {\bf 80}, 692 (1998). 
\bibitem{hd} X.P. Huang, C.F. Driscoll, Phys. Rev. Lett. {\bf 72}, 2187 (1994); H. Brands, P.H. Chavanis, R. Pasmanter, J. Sommeria, Phys. Fluids {\bf 11}, 3465 (1999).
\bibitem{tsallis} V. Latora, A. Rapisarda, C. Tsallis, Phys. Rev. E {\bf 64}, 056134 (2001). 
\bibitem{califano} A. Antoniazzi, F. Califano, D. Fanelli, S. Ruffo,
Phys. Rev. Lett., {\bf 98}, 150602 (2007).
\bibitem{LyndenBell67} D. Lynden-Bell, Mon. Not. R. Astron. Soc. {\bf 136}, 101 (1967).
\bibitem{Chavanis96} P.H. Chavanis, J. Sommeria, R. Robert, ApJ \textbf{471}, 385
(1996); P.H. Chavanis, Ph. D Thesis, ENS Lyon (1996).
\bibitem{super} P.H. Chavanis, Physica A {\bf 359}, 177 (2006).
\bibitem{ellis}  R. Ellis, K. Haven, B. Turkington, Nonlinearity {\bf 15}, 239 (2002).
\bibitem{aaantonov}   P.H. Chavanis, A\&A {\bf  451}, 109 (2006).
\bibitem{barre-06} Y.Y. Yamaguchi, J. Barr\'e, F. Bouchet, 
T. Dauxois, S. Ruffo, Physica A {\bf 337}, 36 (2004).
\bibitem{lifetimeantoni}  M. Antoni, S. Ruffo, A. Torcini,  Europhys. Lett. {\bf 66}, 645 (2004).
\bibitem{lifetime}  P.H. Chavanis, A\&A {\bf 432}, 117 (2005).
\bibitem{morelli}  A. Campa, A. Giansanti, G. Morelli, Phys. Rev. E {\bf 76}, 041117 (2007).
\bibitem{next05}  P.H. Chavanis, Physica A {\bf 365}, 102 (2006).
\bibitem{reentrant} A. W. Francis, {\it Liquid-liquid equilibrium}, (Interscience, NY, 1963);
C. M. Sorensen, Chem. Phys. Lett., {\bf 117}, 606 (1985).

\end{thebibliography}
\end{document}